\documentclass{aa}
\usepackage{rotate,psfig}

\begin{document}

 \thesaurus{03(11.01.1;          
               11.14.1;          
               11.09.1:~3C236)}  

\title{VLBI, MERLIN and HST observations of the giant radio galaxy 3C236}
\author{
R.T.~Schilizzi\inst{1,2} \and 
W.W.~Tian\inst{3,4}      \and
J.E.~Conway\inst{5}      \and
R.~Nan\inst{3}           \and
G.K.~Miley\inst{2}       \and
P.D.~Barthel\inst{6}     \and
M.~Normandeau\inst{7}    \and
D.~Dallacasa\inst{8,9}     \and
L.I.~Gurvits\inst{1}
}
\offprints{R.T.~Schilizzi, {\sl schilizzi@jive.nl}}

 \institute
{
Joint Institute for VLBI in Europe, P.O. Box 2, 7990 AA Dwingeloo, The
Netherlands
\and
Leiden Observatory, P. O. Box 9513, 2300 RA Leiden, The Netherlands  
\and
Beijing Astronomical Observatory and Astrophysics Center of the National Astronomical
Observatories, CAS, 20 Datun Road, Chaoyang Beijing, 100012 PR China 
\and
Max-Planck-Institut f\"ur Radioastronomie, Auf dem H\"ugel 69, D-53121
Bonn, Germany
\and
Onsala Space Observatory, S-439 92 Onsala, Sweden
\and
Kapteyn Institute, P. O. Box 800, 9700 AV Groningen, The Netherlands
\and
Astronomy Department, University of California, Berkeley, CA 94720-3411
USA
\and
Department of Astronomy, University of Bologna, Via Ranzani 1,40127 Bologna, 
Italy
\and
Istituto di Radioastronomia, Via Gobetti 101, 40126 Bologna, Italy
}

\date{Received  9 August 2000; accepted 21 November 2000}

\titlerunning{VLBI, MERLIN and HST observations of 3C236}
\authorrunning{R.T. Schilizzi et al.}
\maketitle

\begin{abstract}

We present VLBI and MERLIN data at 1.66 and 4.99 GHz on the central
component coincident with the nucleus of the giant radio galaxy,
3C236. The nuclear radio structure is composed of two complexes of
emission which are resolved on scales from 1 milli-arcsec (mas) to 1
arcsec. Oscillations with an amplitude of $\sim$ 5$\degr$ can be seen
in the compact radio structure. Spectral index distributions are
plotted at angular resolutions of  10 and 25 mas and allow us to
identify the core component in the south-east emission complex.
Re-examination of the HST WFPC-2 image of 3C236 by de Koff et al.
\cite{dekoff}, shows that the normal to the dust disk in the nucleus is
$\sim$ 30$\degr$ from the plane of the sky and within 12$\degr$ of
parallel to the overall orientation of the radio source. We suggest
that the radio axis is also at an angle of $\sim$ 30$\degr$ to the
plane of the sky and that the north-west jet is on the approaching
side.  This orientation implies an overall size of 4.5 Mpc ($H_{\circ}
= 75$~km\,s$^{-1}$\,Mpc$^{-1}$, $q_{\circ} = 0.5$) for 3C236.  The
coincidence of a dust feature and the south-east compact jet, within
the astrometric errors, leads us to suggest that the dust may be in the
form of a cloud encountered by the jet in the first $\sim$ 400 pc of
its journey out from the nucleus. One-sided emission at 5 GHz on 1~mas
scales would suggest that the jets are ejected initially at $\le$
35$\degr$ to the line of sight, but this is difficult to reconcile with
the obvious orientation stability of the jet system as a whole.
Free-free absorption of the counter-jet may be an alternative
explanation for the one-sideness. At the resolution of WSRT data at 327
MHz, the jet to the south-east is apparently continuous over a distance
of 2.5 Mpc, making this the largest jet known in the universe. It is
likely, however, that activity in the nucleus of 3C236 is episodic but
with a shorter duty cycle than in the double-double sources studied by
Schoenmakers et al.  (\cite{schoenmakers}) and Kaiser et al.
(\cite{kaiser}).


\end{abstract}

\section{Introduction}

Extragalactic radio sources have been studied for many years, but it is
still unclear how they are formed and how they evolve. A crucial
element in the study of their evolution is to identify the young
counterparts of `old' Fanaroff-Riley (FR) class 1 and 2 extended
objects. Recent analyses by Fanti et al. (\cite{fanti}), Readhead et al
(\cite{readhead}), and Snellen et al. (\cite{snellen}) suggest that
good candidates for young radio sources are the Gigahertz Peaked
Spectrum (GPS) and Compact Steep Spectrum (CSS) sources with symmetric
morphologies. They are small in angular size as expected for young
sources, and could plausibly expand to larger sizes while decreasing in
luminosity as required to match the statistics of the numbers of
compact and extended objects. The propagation velocities of hot-spots
in some small size symmetric sources have been measured at a few tenths
of the speed of light by Owsianik \& Conway (\cite{owsianik}) and
Tschager et al. (\cite{tschager}); these velocities indicate dynamic
ages of a few hundred to a few thousand years. In a few objects we see
both a GPS or CSS component associated with the nucleus of the galaxy
and extended FR1 or FR2 lobes either side of the nucleus. These sources
with recurrent activity provide us with case studies of the evolution
of individual objects which constrain models for the evolution of radio
sources in general.  One such object is 3C236.

3C236 is the largest radio source known in the universe and this in
itself makes the source a powerful laboratory for the study of radio
source evolution. The remarkable agreement between the orientation of
the inner and outer radio structure was one of the original arguments
that a massive black hole was the source of the AGN energy. Its angular
extent of 39 arcmin corresponds to a projected size of 3.9 Mpc at the
redshift of its associated 17$^m$ galaxy ($z=0.1005 \pm 0.0005$, Hill
et al. \cite{hill}, $H_{\circ} = 75$~km\,s$^{-1}$\,Mpc$^{-1}$,
$q_{\circ} = 0.5$). Its total radio luminosity of $1.5\times
10^{43}$~erg\,s$^{-1}$ (Mack et al. \cite{mack2}) puts it in the
transition zone between FR1 and FR2 systems. Its morphology is also
intermediate between FR1 and FR2 on large scales (Strom et al.
\cite{strom}, Barthel et al.  \cite{barthel}) with a component
coincident with the nucleus of the galaxy and two lobes of quite
different morphology, one (SE) narrow and edge-brightened (FR2-like)
and clearly connected to the nucleus by a low surface brightness jet
visible in WSRT 92 cm data (Mack et al. \cite{mack1}), and the other
(NW) more diffuse and centre-brightened (FR1-like) and extended towards
the nucleus but not obviously by a narrow jet.  Spectral index studies
at wavelengths from 92 to 2.8 cm by Mack et al.  (\cite{mack2}) show
that the spectral index in the NW and SE lobes gradually steepens from
the edge of each lobe back towards the nucleus; in addition, the
spectral index in the NW lobe is generally steeper than in the SE
lobe.

About two-thirds of the total luminosity originates in the steep
spectrum ($\alpha = -0.6$, $S \propto \nu^{\alpha}$) nuclear
component.  Earlier high resolution studies (Fomalont \& Miley
\cite{fomalont}; Schilizzi et al. \cite{schilizzi1}) have shown its
overall size to be 1.3 arcsec corresponding to 2.2 kpc, with an
asymmetric morphology (Barthel et al. \cite{barthel}) which is similar
in some respects to that on the larger scale.

An HI absorption line has also been detected towards the nucleus of
3C236 by van Gorkom et al. (\cite{vgorkum}) with the VLA and mapped in
detail with the VLBA by Conway \& Schilizzi (\cite{conway}). The high
angular resolution results show that most of the absorption takes place
about 400 pc from the nucleus at the tip of the SE jet, and is possibly
due to a jet-cloud interaction.

HST/WFPC-2 observations of 3C236 by de Koff et al. (\cite{dekoff})
revealed the presence of a dust ring in the nucleus $\sim 10$~kpc in
diameter and oriented perpendicular to the overall radio axis. In
addition, a dust lane was observed on scales of 1 kpc seemingly
connected to the dust disk by a strand of dust going from one to the
other. 3C236 is unusually bright at FIR wavelengths (Hes et al.
\cite{hes}, Willott et al. \cite{willott}) which implies that there is
warm dust in the interstellar medium, presumably the dust seen at
optical wavelengths. Recent paper by O'Dea et al. (\cite{o'dea})
presents HST STIS/MAMA near-UV observations of the nucleus of 3C236.
Four star-forming regions are found in an arc along the edge of the
dust lane.

The age of 3C236 is quite uncertain. An analysis of the energy and
magnetic field content of the large scale lobes by Mack et al.
(\cite{mack2}) led to a maximum age for particles in the source of
$2\times 10^7$~years (for the hotspot at the end of the SE lobe).  On
the other hand, assuming an average propagation speed for the hotspot
in the SE lobe of $0.03c$ (Lacey \& Cole \cite{lacey}, Scheuer
\cite{scheuer}), the age of 3C236 is $2\times 10^8$~years.
Reacceleration at the hotspot or resupply of electrons to the hotspot
via a jet is therefore necessary unless the speed of propagation is an
order of magnitude greater than $0.03c$. The age of the CSS nuclear
component is equally uncertain, from a few 10$^3$ years to a few
$10^4$~years old, depending on the speed of propagation. The estimated
ages of the star-forming regions found by O'Dea et al. (\cite{o'dea})
range from $\sim 10^{7}$ to $10^{8}-10^{9}$~years.

In this paper we report on an extensive study of the nuclear component
in 3C236 using combinations of VLBI and MERLIN data at 18 and 6 cm.
Spectral index distributions of the various components on angular
scales of 10 mas are presented and the physical characteristics of the
nuclear emission are discussed. We re-examine the HST data by de Koff
et al. (\cite{dekoff}) and its relation to the radio structure on both
the large and small scales.

\section{ Observations and data reduction } 
\subsection{ VLBI data } 

Global VLBI observations of 3C236 were carried out at 1.663 GHz in
September 1984 using telescopes at 16 locations: Jodrell Bank (25~m Mk2
telescope), Cambridge~(15~m), Defford~(25~m), Westerbork-1 (one 25~m
telescope from the array), Effelsberg~(100~m), Onsala~(25~m),
Simeiz~(22~m), Iowa~(18~m), Haystack~ (36~m), Maryland Point~(25~m),
Green Bank~(43~m), Penticton~(26~m), VLA-27 (twenty seven 25~m
telescopes), Fort Davis~(25~m), Owens Valley~(40~m), and Hat
Creek~(25~m).

\begin{figure}[h]
\centerline{
{
\psfig{figure=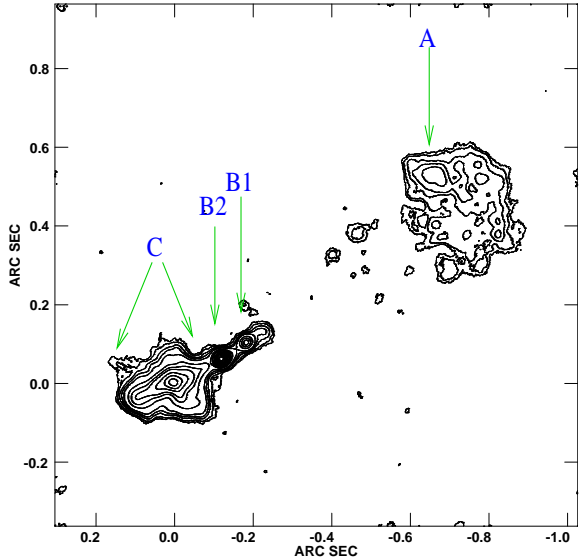,width=8.0cm,bbllx=52pt,bblly=179pt,bburx=519pt,bbury=658pt,clip=
}}
}
\caption{The kilo-parsec scale central component of 3C236 at 1.663 GHz;
restoring beam is a circular Gaussian  25 mas FWHM; contour levels are
$-$1, 1, 1.5, 2.5, 3.5, 5, 10, 15, 20, 30, 40, 50, 60, 75, 90\%  of the
peak brightness, which is 251 mJy/beam. Component B2 is the nucleus.}
\label{Fig. 1}
\end{figure}

\begin{figure}[h]
\centerline{
{
\psfig{figure=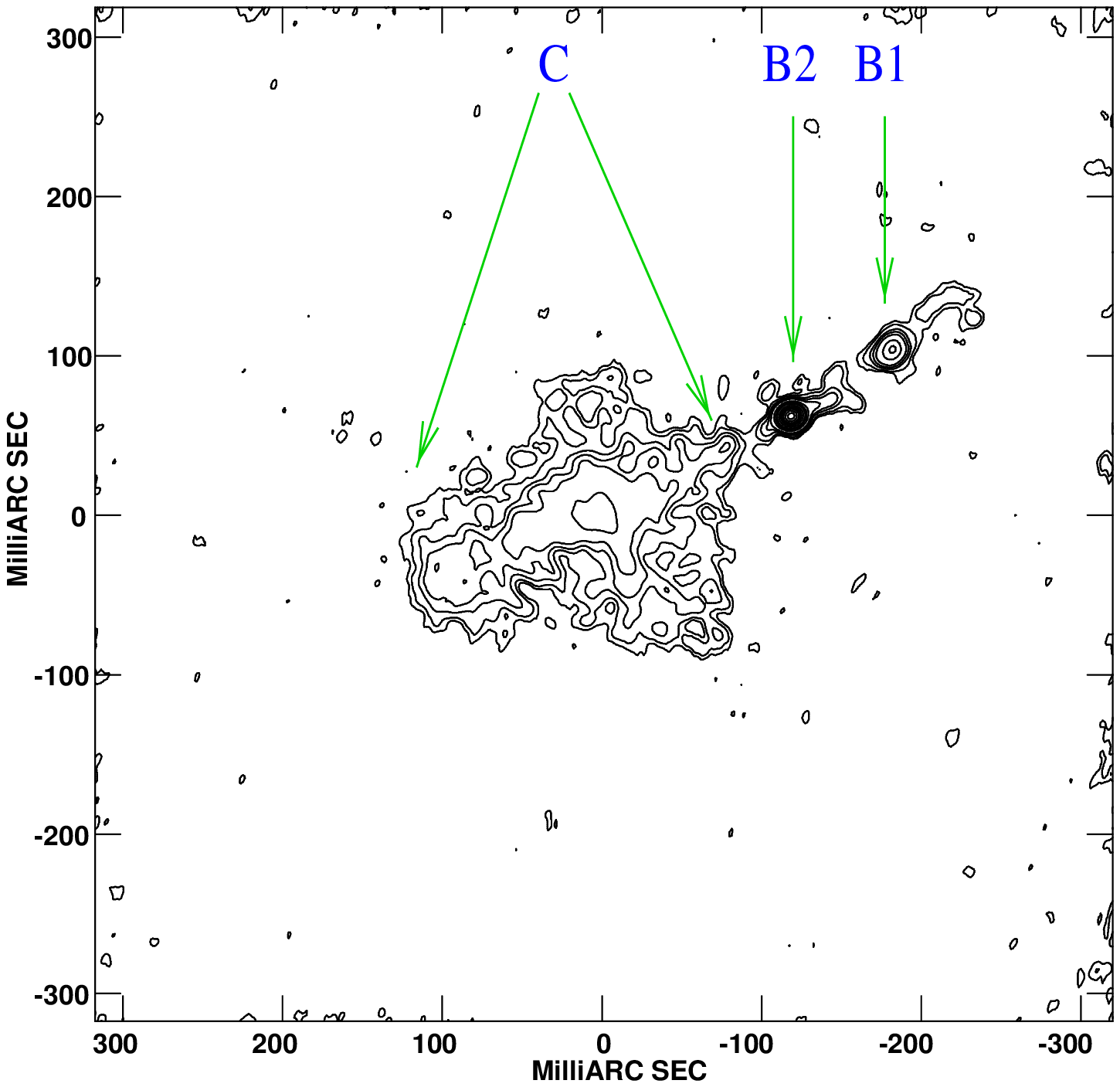,width=8.0cm,bbllx=52pt,bblly=188pt,bburx=508pt,bbury=639pt,clip=
}}
}
\caption{Components B1-B2-C: the central 650 pc of the nucleus of 3C236
at 1.663~GHz. The restoring beam is a circular Guassian 10 mas FWHM;
contour levels are $-$1, 1, 1.5, 2.5, 3.5, 5, 10, 15, 20, 30, 40, 50,
60, 75, 90\% of the peak brightness, which is 216 mJy/beam. Component
B2 is the nucleus.}
\label{Fig. 2}
\end{figure}

Additional global observations at 4.987~GHz were carried out in
September 1989 using telescopes at 9 locations: Onsala~(25~m),
Westerbork-14 (fourteen 25~m telescopes), Jodrell Bank (25~m Mk2
telescope), Effelsberg~(100~m), Haystack~(36~m), Green Bank~(43~m),
VLA-27, Pie Town~(25~m), and Owens Valley~(40~m). Data for the
1.663~GHz observations were recorded with Mk2 equipment (1.8 MHz
bandwidth) and correlated on the CIT/JPL data processor at the
California Institute of Technology. The 4.987~GHz data were recorded
with Mk3 equipment in mode~B (28~MHz bandwidth) and correlated at the
Max-Planck-Institut f\"ur Radioastronomie in Bonn.

\subsection{MERLIN data} 

An important addition to the aperture plane coverage for 3C236 was made
by adding MERLIN data obtained at the same frequency but at a different
epoch to the VLBI data. Observations at 1.658 GHz were made with seven
MERLIN telescopes in September 1993, and at 4.996 GHz with six
telescopes in September 1991. In combining the VLBI and MERLIN
datasets, we assume that the structure does not vary between the epochs
of observation, and that the small scaling errors caused by combination
of data at slightly different frequencies is acceptable. In particular,
we assume that the 1-2\% variations in the flux density of the central
compact component seen at 5 GHz over a period of 8 years (see sect.
3.2) have a negligible effect on the matching of the MERLIN and VLBI
data taken 2 years apart. The phase centres of the VLBI and MERLIN
dataset were originally matched (Schilizzi et al. \cite{schilizzi2}) by
assuming that the peak emission in the MERLIN image coincided with
component B2 in the VLBI data (see sect. 3). There were no common
baselines to verify this assumption.  However, the VLBA image at
1292~MHz of Conway \& Schilizzi (\cite{conway}) showed that this
assumption is incorrect and that the phase centre of the MERLIN data
should coincide with the centre of component C in the VLBI data. This
has been rectified in the data reduction reported here.

The data were reduced using standard procedures in the AIPS package.
Images of varying resolution were made by applying tapers to the data
in the $uv$-plane.

\section{Results} 
\subsection{The structure of 3C236 at 1.663 GHz} 

Fig.~\ref{Fig. 1} shows the 2 kpc radio structure in the nucleus of the
galaxy with an angular resolution of 25 mas, and Fig.~\ref{Fig. 2} the
structure in the central 650 pc with an angular resolution of 10 mas.
The visibility data derived from the images shown are a good fit, within
the errors, to the observed data on all baselines.

\begin{figure*}
\vspace{30mm}
\begin{picture}(40,100)
\put(-31,-60){\includegraphics{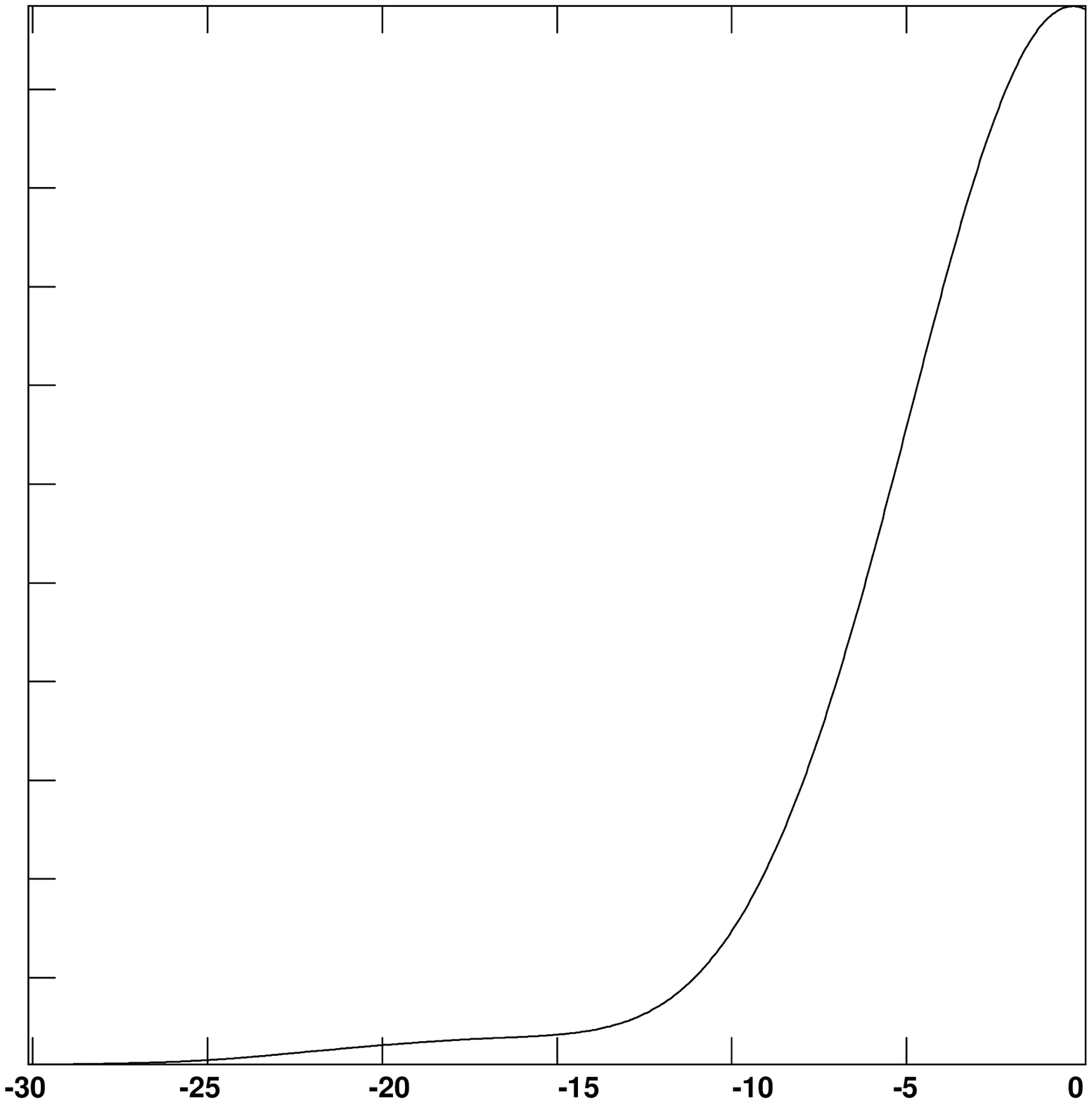}}
\put(127,-60){\includegraphics{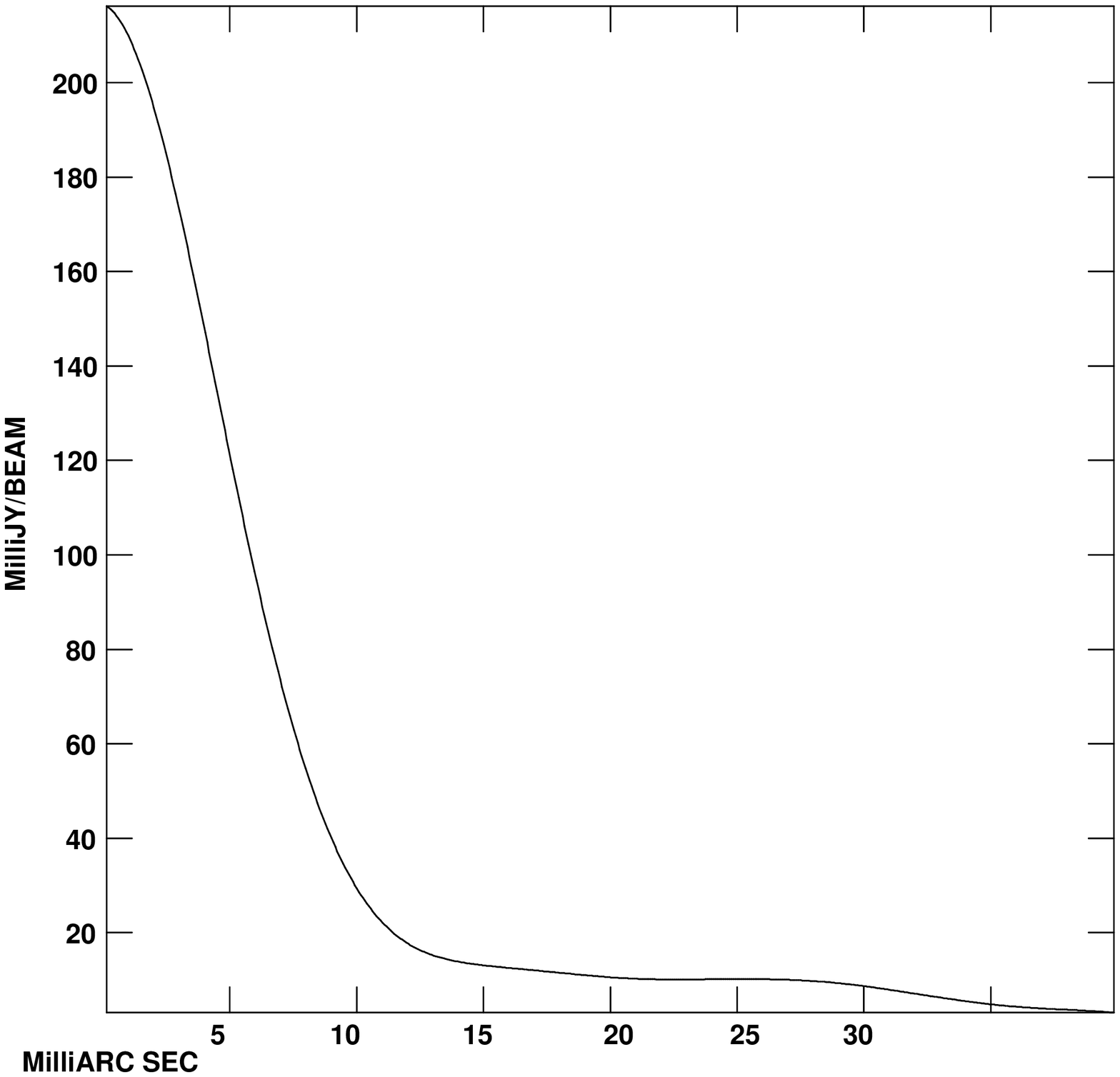}}
\end{picture}
\caption{A slice along the ridge of jet emission seen either side of
component B2 at 1.663 GHz in Fig. 2. The profile shows that the jet brightness is
very similar on either side of B2 as expected for a source near the
plane of the sky }
\label{Fig. 3}
\end{figure*}

Fig.~\ref{Fig. 1} shows two main emission areas, the central strong
emission region (B1--B2--C) and the extended radio lobe in the
northwest (A) which appears to be a diffuse plateau structure with low
surface brightness.  Components B1, B2, and C are more clearly defined
in Fig.~\ref{Fig. 2}. The brightest and most compact component, B2, was
earlier identified as the location of the nucleus of 3C236 on the basis
of its spectral index and compactness (Schilizzi et al.
\cite{schilizzi2}). Component B1 is a slightly resolved knot in the jet
to the NW, and component C appears to be a lobe with a bright spine
which defines the radio jet to the SE. The overall radio axis defining
the long term orientation of the source is given by the line joining
the prominent features in the large scale extended lobes and passing
through the nuclear component; this has a position angle (P.A.) of
$123\degr \pm 2\degr$ (Schoenmakers et al. \cite{schoenmakers}, see
also Fig.~\ref{Fig. 9}). Not all the individual components in the 2 kpc
radio structure lie on this line. One that does is the slightly
resolved component B1 which lies 77 mas north-west of  B2 in P.A.
$-56\fdg 5 \pm 0\fdg 5$. In contrast, the main ridge line of the
lobe-like component C on the other side of B2, lies in P.A.~$116\degr
\pm 2\degr$. The most prominent feature in component A is about 720 mas
from B2 and located north of the overall source axis. Weak emission
features are seen between this feature and the B1--B2--C complex which
probably delineate the path of the jet to the north-west. Closer to the
nucleus, B2, the jets appear to oscillate in direction; the initial
direction of the jets emanating from B2 to the northwest and south-east
is in P.A. $114\degr \pm 0\fdg 5$ before changing smoothly to a P.A.~of
$122\degr$ about 50 mas from B2 and back again to $114\degr$ mid-way
through component C. These observations suggest that the jets emanating
from B2 oscillate in direction within a cone of half angle $\sim 5\degr
- 7\degr$. Note that the brightness of the jets close to B2 are very
similar (Fig.~\ref{Fig. 3}). The transverse structure in component
C which lies 137 mas south east of B2 may be evidence of a shock
propagating back through the lobe emission (Barthel et al.
\cite{barthel}, Schilizzi et al. \cite{schilizzi2}, Conway and
Schilizzi \cite{conway}).

\begin{figure}[h]
\centerline{
{
\psfig{figure=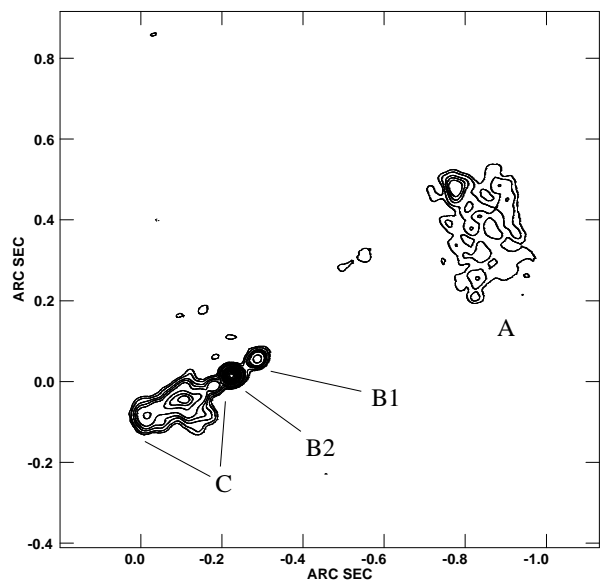,width=8.0cm,bbllx=36pt,bblly=141pt,bburx=576pt,bbury=688pt,clip=
}}
}
\caption{The kilo-parsec scale central component of 3C236 at 5~GHz.
The restoring beam is a circular Gaussian 25 mas FWHM;
contour levels are $-$1, 1, 1.5, 2.5, 3.5, 5, 10, 15, 20, 30, 40, 50,
60, 75, 90\% of peak brightness, which is 272~mJy/beam.}
\label{Fig. 4}
\end{figure}

\begin{figure}[h]
\centerline{
{
\psfig{figure=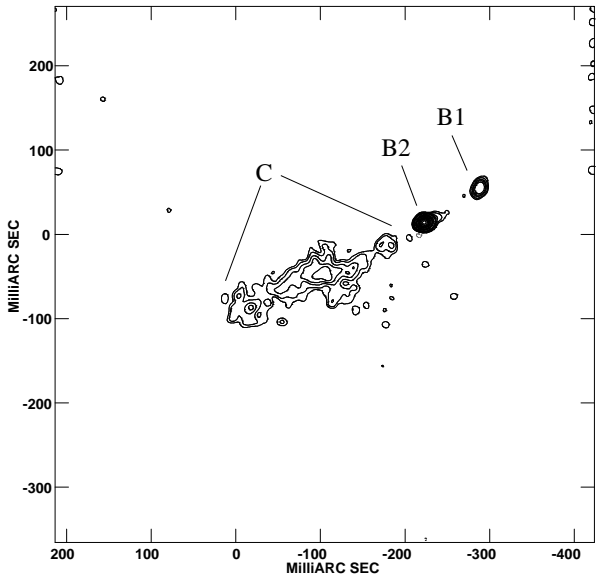,width=8.0cm,bbllx=36pt,bblly=130pt,bburx=576pt,bbury=688pt,clip=
}}
}
\caption{Components B1--B2--C; the central 650 pc of the nucleus of
3C236 at 5~GHz; the restoring beam is a circular Gaussian 10 mas FWHM;
contour levels are $-$1, 1, 1.5, 2.5, 3.5, 5, 10, 15, 20, 30, 40, 50,
60, 75, 90\% of peak brightness, which is 232~mJy/beam.}
\label{Fig. 5}
\end{figure}

\begin{figure}[h]
\centerline{
\psfig{figure=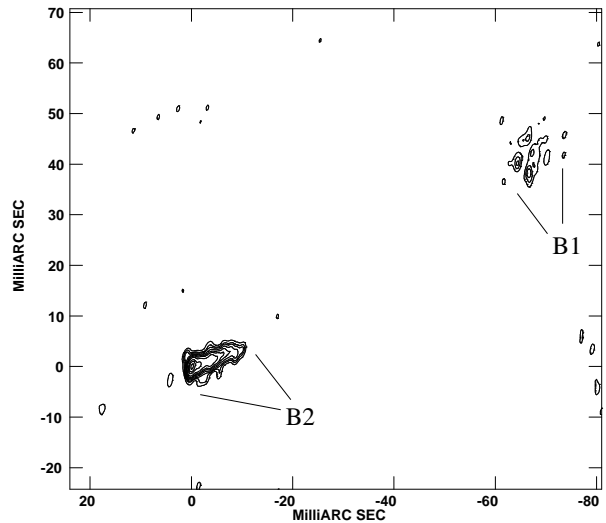,width=8.0cm,bbllx=36pt,bblly=161pt,bburx=576pt,bbury=658pt,clip=
}}
\caption{The central 150 pc of 3C236 at 5 GHz; component B1 is to the
northwest and component B2 (containing the nucleus) is to the
southeast. The restoring beam is 1.5 mas; contour levels are 1, 2, 3,
8, 12, 20, 40, 60, 90\% of peak brightness, which is 85.54~mJy/beam.}
\label{Fig. 6}
\end{figure}

\begin{table*}
\label{table 1}
\caption{Component parameters of 3C236}
\begin{center}
\begin{tabular}{cccccccc}
\hline \hline
Freq   & S$_{tot}$ & Component &   S   &  Sep  &  P.A. & w1$\times$w2 &  P.A. 
\\[0.1cm] 
(GHz)  & (Jy)      &           & (mJy) & (mas) & (deg) & (mas)        & (deg) 
\\[0.1cm] 
\hline
1.663  & 2.8       & B2        &  252  & --    &  --   & 12$\times$9  &  106  
\\[0.1cm]
       &           & B1        &   64  & 77    & $-$58 & 17$\times$14 &  142  
\\[0.1cm] 
       &           & C         & 1560  & 135   & 119   &  --          &   --  
\\[0.1cm] 
       &           & A         & 524   & 700    & -58    &  280$\times$280          
&   --  \\[0.2cm] 
 \hline
4.987  & 1.4       & B2        & 267   & --    &  --   & 12$\times$9  &  108  
\\[0.1cm] 
       &           & B1        & 35    & 75    & $-$58 & 16$\times$12 &  156  
\\[0.1cm] 
       &           & C         & 500   & 134   & 118   &  --          &   --  
\\[0.1cm] 
       &           & A         & 220   & 720    &  -58   &  280$\times$210          
&   20  \\[0.1cm]  
\hline
\hline
\\[1cm]
\end{tabular}
\end{center}
\end{table*}

\subsection{The structure of 3C236 at 4.987 GHz} 

Fig.~\ref{Fig. 4} shows the central radio structure of 3C236 at 4.987
GHz at an angular resolution of 25 mas, and Fig.~\ref{Fig. 5} the
central 650 pc at a resolution of 10 mas. The main features at 4.987
GHz match those at 1.663 GHz including the weak feature in the north of
component A. The surface brightness sensitivity at 4.987 GHz is not
sufficient to recover the low level emission in the outer regions of
component C seen in Fig.s 1 and 2, or to reliably recover the details
of the low level emission in component A.  The double-sided jet
structure associated with B2 at 1.663 GHz is apparently one-sided at
4.987 GHz pointing northwest in the same direction as Fig ~\ref{Fig.
2}. The oscillating behaviour of the jet in component C is clearly
visible within a cone of half angle 5$\degr$ centred on B2.

When the full angular resolution of the VLBI data is used (1.5~mas,
uniform weighting), a different picture emerges (Fig.~\ref{Fig. 6}).
The centre of component B2 is revealed as a multi-component structure
in position angle $112\degr$, with the brightest component at the
eastern end. The presence of component B1 is also visible in the data
although it is well resolved. This is the highest angular resolution
data on the nucleus itself, and it is remarkable that at this
resolution, the structure appears to be a one-sided jet whereas, on
larger scales, the structure is two-sided.

Comparison of visibility amplitudes of these 5 GHz data with the
corresponding data on 5 baselines from April 1981 (Barthel et al.
\cite{barthel}), shows evidence for brightening of the most compact
component in the structure by a few tens of mJy over the 8 year period.
Comparison of the apparent positions of the minima in the visibility
data shows no consistent evidence for change in the positions of
individual components at a level of 0.065 mas/year which corresponds to
an upper limit on the speed of separation of $0.4c$.

\subsection{Source component parameters}

Table 1 gives information on  the source parameters measured at 1.663
and 4.987 GHz.  1) observing frequency; 2) total flux density;   3)
component name;  4) flux density of the component; 5) distance from B2;
6) P.A. of the component with respect to B2; 7) mean deconvolved
component sizes at half power, w1 and w2, determined from
two-dimensional Gaussian fits for the components that could be
approximated well by a Gaussian; 8) P.A. of the component in degrees.
The flux densities in columns 2 and 4 were measured from the images
using the AIPS task TVSTAT, and have typical errors of 5\%. The
separations and position angles in columns 5 and 6 were measured from
the images by hand and have typical errors of 5\% and $1\degr$
respectively. The sizes and P.A.s of individual components in columns 7
and 8 were determined by 2-dimensional gaussian fits where possible and
have typical errors of 10\% and $5\degr$ respectively.

\subsection{The spectral index distribution between 1.663 and 4.987 GHz}

We have constructed spectral index distributions for 3C236 at angular
resolutions of 10 and 25 mas on the assumption that the source
structure did not vary during the time spanned by the observations
reported here. There is no evidence for flux density variations greater
than a few tens of mJy (see sect. 3.2) at either frequency. The images
at 1.663 and 4.987 GHz were convolved to the same resolution, then
super-imposed assuming the position of the peak brightness in component
B2 is the same at each frequency. Spectral indices  were computed for
each pixel for which the intensities in the two images were both
greater than 8$\sigma$.

\begin{figure}[h]
\centerline{
\rotate[r]
{
\psfig{figure=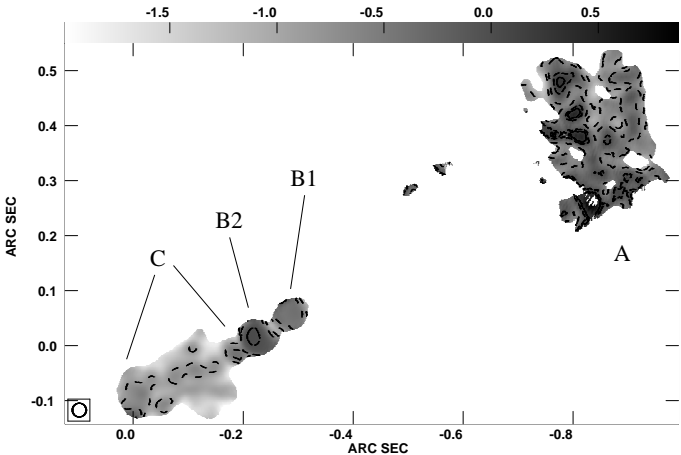,width=7.0cm,bbllx=36pt,bblly=66pt,bburx=543pt,bbury=726pt,clip=
}}
}
\caption{The spectral index distribution across the central components
of 3C236, at an angular resolution of 25 mas, for the frequency range
from 4.987 to 1.663~GHz. The greyscale range extends from $-$2.0
(faintest grey) to 0.9 (black); contours are shown for spectral indices
of $-$0.85, $-$0.5, 0, 0.2}
\label{Fig. 7}
\end{figure}

\begin{figure}[h]
\centerline{
\rotate[r]
{
\psfig{figure=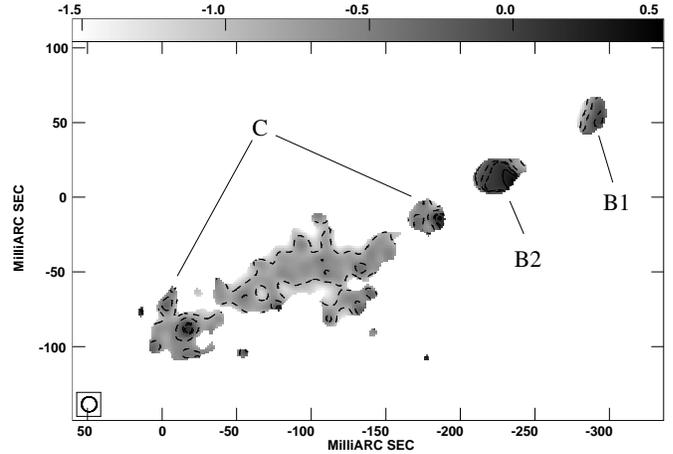,width=7.0cm,bbllx=36pt,bblly=76pt,bburx=543pt,bbury=716pt,clip=
}}
}
\caption{The spectral index distribution across the central 650 pc of
3C236 at an angular resolution of 10 mas, for the frequency range from
4.987 to 1.663~GHz. The greyscale range extends from $-$1.5 (faintest
grey) to 0.5 (black); contours are shown for spectral indices of
$-$0.9, $-$0.5, $-$0.1, 0, 0.2}
\label{Fig. 8}
\end{figure}

Fig.~\ref{Fig. 7} shows the overall spectral index distribution
across the source at an angular resolution of 25 mas.  The spectrum is
steep ($-1.5 \le \alpha \le -0.5$) over most of the structure in the
central 2 kpc with the exception of component B2 which has a flat
spectral index ($\alpha = 0.05$) and the warmspot in the north of
component A which appears to have an inverted spectrum ($\alpha =
0.3$). The other regions of apparent inverted spectrum in component A
may not be reliable in view of the substantial difference in
$uv$-coverage at the two frequencies; these differences are capable of
causing ambiguities in the reconstruction of the total intensity of
this very extended component.  It is interesting to note that the
oscillating jet to the SE appears to have a flatter spectral index
where it passes through the centre of component C.

A more detailed picture of the distribution of spectral index in the
central 500~pc is shown in Fig. ~\ref{Fig. 8}. The spectral index in B2
varies from $-0.5$ in the east to 0.4 at the base of jet to the
northwest; further out along the jet it steepens again reaching
$-1.0$.  There is also a gradation of spectral index in component B1
from east to west, changing from $-1.0$ to 0. As remarked earlier, the
ridge in component C has a spectral index of about $-0.5$ in the
centre, which falls off to $-1.0$ towards the edges. At the location of
the transverse feature in component C, the spectral index appears to
flatten from $-0.9$ to $-0.5$.  At the eastern end of C, the spectral
index of the peak seen in the total intensity maps has a flat spectral
index of about zero.  From this point to the leading edge of the jet is
the region in which Conway \& Schilizzi (\cite{conway}) detect maximum
HI opacity.

\section{Discussion}
\subsection{A re-examination of the HST data}

\begin{figure}[h]
\centerline{
{
\psfig{figure=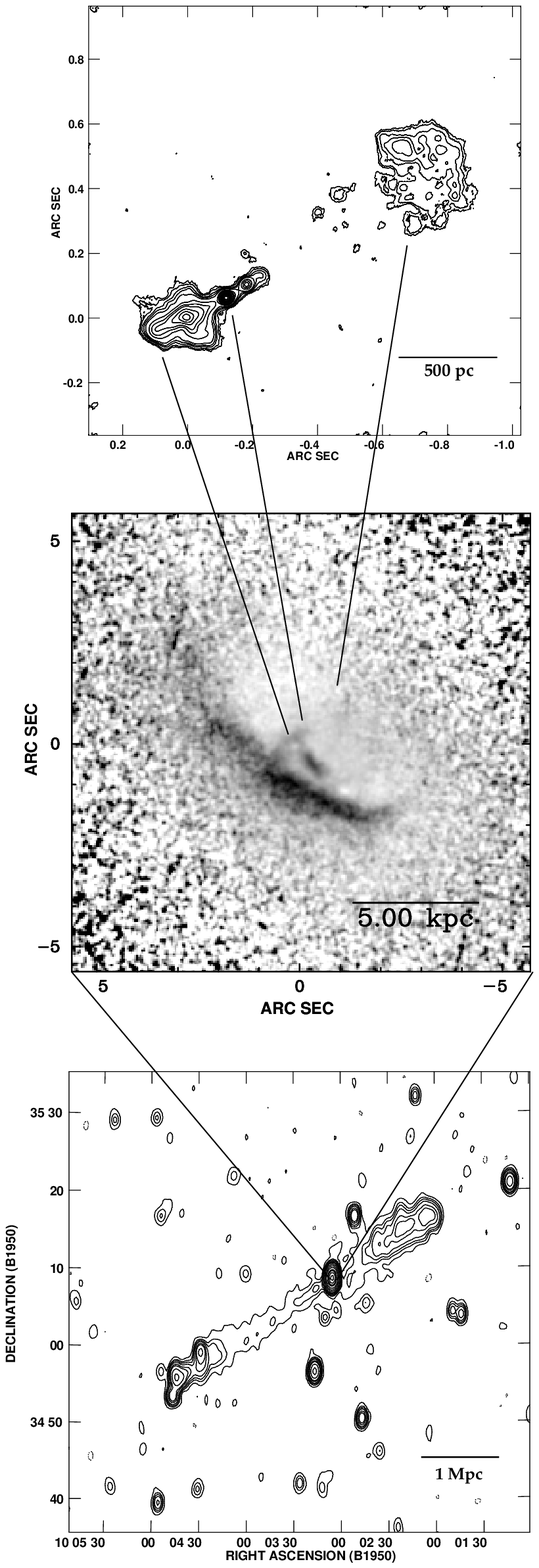,width=6.0cm,bbllx=184pt,bblly=56pt,bburx=422pt,bbury=772pt,clip=
}}
}
\caption{ The images of 3C236 from observations with VLBI at 18 cm
(top, see Fig. 1 for details), HST absorption at 7000 \AA~(middle), and
the WSRT at 92 cm (bottom, Schoenmakers et al. \cite{schoenmakers}).
For the WSRT map, the restoring beam is 45 arcsec; contour levels
$-$0.1, 0.1, 0.3, 0.5, 0.8, 1.4, 2.4, 8, 15, 30, 40, 60, 90\% of peak
brightness, which is 7.24~Jy/beam. The optical nucleus is at (0,0) in
the middle panel.}
\label{Fig. 9}
\end{figure}

The HST snapshot of 3C236 made with the WFPC-2 using a red broad-band
filter centred near 7000~\AA,~shows clear evidence of a substantial
dust ring (de Koff et al.  \cite{dekoff}). Fig.~\ref{Fig. 9} (middle)
reproduces the absorption model for the galaxy derived by de Koff et
al., in which the main component of dust is in the form of a ring of
radius $\sim$ 5 kpc whose apparent symmetry axis differs in position
angle by 15 - 20 $\degr$ from  the overall radio axis defined by the
outer edges of the large scale structure (Fig. ~\ref{Fig. 9}, bottom).
The symmetry axis of the ring is aligned with the apparent minor axis
of the galaxy itself, within the errors. A dust feature $\sim$1.5 kpc
long lies internal to the ring and approximately parallel to it; this
may be the remnant of another ring.  A further faint dust feature can
also be seen emanating from the nuclear region towards the SE.

From the ellipticity of the disk, we derive an apparent inclination
angle of the radio source to the line of sight of $\sim 60\degr$
considering that the radio jets are approximately perpendicular to the
dust ring. Kotanyi \& Ekers (\cite{kotanyi}) first pointed out that
large scale radio jets are roughly perpendicular to dust disks and
lanes, and this has been amply confirmed and extended to the nuclear
regions by de Koff et al. (\cite{dekoff}), in particular for well
defined disks and lanes. It is interesting to note that in NGC~4261,
Jaffe et al. (\cite{jaffe}) found a very similar offset in P.A. between
the radio jet and the symmetry axis of the dust disk in that galaxy.
These authors refer to Rees (\cite{rees}) in pointing out that
Lense-Thirring precession will force the part of the accretion disk
nearest to the black hole to be axi-symmetric about the rotation axis
of the hole, whatever its spin axis at large radius. At small distances
from the black hole the jet and disk axes should coincide. In the case
of 3C236 the linearity of the large scale radio jet to the south-east
implies that the black hole axis has been oriented at 122 $\degr$ for
the lifetime of the source, and that the dust ring axis is offset from
this on kpc scales. This offset could arise if the dust has been
captured from a smaller galaxy which has been cannibalised by 3C236.
The oscillation in direction of the compact and large scale jets (see
next section), if due to oscillation in the black hole axis, may well
be related to a merger event. The mass of dust in the ring estimated by
de Koff et al. (\cite{dekoff}) of $3\times 10^6 M_{\sun}$ seems
unlikely by itself to cause the oscillations through torque on the
accretion disk.

The inclination of the radio axis in 3C236 implies that the northwest
jet is approaching and  the SE jet receding. It also implies that the
true size of 3C236 is 4.5 Mpc rather than 3.9~Mpc calculated for an
inclination angle of $90\degr$.

The nominal centre of the galaxy in the optical absorption image lies
about 0.2~arcsec south of the northwest end of the small dust feature
which extends $\sim 500$~pc in P.A.~$110\degr - 115\degr$ before
apparently connecting into another dust strand at P.A.~$\sim$145
$\degr$.  Since the astrometric accuracy of the HST images is $\sim
0.1$~arcsec, it is possible that the position of the nucleus coincides
with the northwest end of this dust feature. Assuming this to be the
case, the fact that the compact radio jet/lobe to the SE (Fig.
~\ref{Fig. 9} top; see also Fig.~\ref{Fig. 1}, component C) has the
same orientation (P.A.~$\sim 116\degr$) and almost the same extent ($\sim
370$~pc) as the feature, suggests that the radio emission and dust may
be related.

We can make some qualitative remarks on this apparent connection. It
seems unlikely that the jet has entrained dust on its way out for the
simple reason that the jet itself is not in contact with the neutral
interstellar medium. It propagates through the relativistic plasma in
the centre of the radio `lobe' which is the shocked plasma from the jet
that has passed through the hotspot. Indeed De Young (\cite{deyoung})
has pointed out that the passage of a strong high-speed shock
associated with a radio jet will destroy dust grains by sputtering
processes on timescales of $10^4 - 10^5$~years, comparable with the age
of component C. The most the radio jet can do is concentrate the dust
and hence increase the column density and obscuration along certain
lines of sight. This could take place via the expansion of the radio
lobe that generates a bow-shock propagating out into the neutral
interstellar medium. Of course the speed of the bowshock should not be
so high that the dust is destroyed.

Perhaps the simplest explanation of the apparent radio-dust connection
is that the dust feature represents a dense cloud that has happened to
wander into the path of the jet as shown by the HI absorption data. In
this case, the optical dust feature traces the mass which shapes the
radio structure, rather than the radio structure generating the optical
structure via entrainment.

\subsection{Radio morphology}

The radio morphology of 3C236 shows structure from less than a few
milli-arcseconds to 39 arcmin, a range of 10$^6$. At first sight, the
small and large scale structures are similar in some respects but not
in others. The components to the NW on the small and large scale have
approximately the same opening angle, 20$\degr$ seen from the nucleus,
and they are both much more resolved in the transverse direction than
the components to the south-east. The structure to the SE is clearly
very well collimated on the large scale, but on the small scale the SE
component flares about 45 mas (74 pc) from the nuclear component. The
higher resolution VLBI data on this component shows that there is still
collimated flow through the centre of the flaring component. The
flaring is most likely related to a backflow/transverse shock resulting
from the putative collision between the jet and a cloud $\sim 225$ mas
(370~pc) from the nucleus (Conway \& Schilizzi \cite{conway}).

The arm-length ratios (the ratio of the distances from the nucleus to
the extremities of the components) of the inner and outer structure are
quite different. The end of the NW component on the small scales is
considerably further from the nucleus than the end of the SE component,
whereas on the large angular scales, the reverse is true. The
arm-length ratio for the large scale structure is $\sim 1.6$, defined
as the distance of the SE hotspot from the nucleus divided by the
distance of the end of the NW structure from the nucleus. This is an
average arm-length ratio for giant radio galaxies (Schoenmakers et al.
\cite{schoenmakers}). Defined in the same way, the arm-length ratio for
the compact structure is $\sim 0.3$. The difference may reflect the
fact that the large scale structure is shaped by the interaction of the
jet with the inter-galactic medium and the small scale structure by the
interaction of the jet with the ISM. It is also intriguing that the Mpc
SE counter-jet is considerably longer and better collimated than the NW
jet. This suggests that the jet to the NW has encountered a denser
galactic or intergalactic medium which has reduced its outward motion.
The large scale arm-length asymmetry cannot be due to light travel time
effects (Longair \& Riley \cite{longair}), in particular since we
conclude that the NW side is approaching.

The fact that 3C236 is so large and has remained so stable in overall
orientation argues that it resides in a relatively empty environment.
AGN in more dense environments (eg. Cygnus A) radiate more of their
energy and/or become Wide Angle Tail sources, whereas more of the
energy in objects like 3C236 goes into outward expansion (eg. Barthel
\& Arnoud \cite{barthel2}).

The 92 cm image of 3C236 (Fig. ~\ref{Fig. 9} bottom) provides strong
evidence that the jet to the SE has been on for most of the life of the
radio source. At the resolution of the measurements (45 arcsec), the
jet appears to originate in the nucleus and continue for $\sim 2.5$~Mpc
before terminating in a double hotspot (Barthel et al. \cite{barthel}).
The jet appears to broaden in the transverse direction as it moves
further away from the nucleus, eventually subtending an angle of $\sim
10\degr$ which is considerably larger than the angle subtended by the
separation of the hotspot at the nucleus, 3$\degr$. Barthel et al.
(\cite{barthel}) show that the ridge line of the SE structure wiggles
over the final 600 kpc of its trajectory, so that the cone of opening
angle 10$\degr$ may well be the envelope of the jet's transverse
motion. The jet ridge line in the 92 cm image is located along the
southern edge of the cone but curves slightly northward before reaching
the double hotspot. On average, the jet and counterjet directions have
remained very stable over the lifetime of the source.

Close to the nucleus, the SE jet is apparently blocked due to a
collision with a cloud (Conway \& Schilizzi \cite{conway}) which has
caused what looks like a transverse reverse shock back towards the
nucleus.  This is presumably of recent origin since the jet must
re-start further out, as we see in the 92 cm data. The highest
resolution data on the large scale jet was published by Barthel et al.
(\cite{barthel}) at 21 cm but does not have sufficient sensitivity to
trace the jet in close to the nucleus.

Superficially, the evidence from the large-scale jet to the SE is that
activity in 3C236 is $\it{continuous}$ rather than $\it{recurrent}$.
However, the 92 cm image in Fig.~\ref{Fig. 9} has a resolution of 45
arcsec or 74 kpc projected on the plane of the sky. Gaps in the
emission of this size would not be detected.  Taking a jet speed of
$0.1c$ (a typical speed for compact symmetric objects on similar scales
- Owsianik \& Conway \cite{owsianik}; Tschager et al. \cite{tschager}),
and assuming that we actually see the jet and not the backflow in a
thin lobe, we would not detect 'off' periods in jet activity lasting
less than $2\times 10^6$~years. This period is longer than the probable
length of the present `on' period represented by the kpc central
component, $10^4 - 10^5$~years. The evidence from the structure to the
NW points more towards episodic rather than continuous activity. In the
central kiloparsec, the jet is traceable via a number of knots on its
way from the nucleus via component B1 out to the warmspot in component
A at a distance of $\sim 1.2$~kpc. Insufficient sensitivity may limit
our ability to detect more of the jet on these sub-galactic scales. On
larger scales, the emission extends $\sim 1.4$~Mpc in what appears to
be a more relaxed jet than its well-collimated counterpart to the SE.
In the 21 cm image made by Barthel et al. (1985) there is a prominent
feature in the NW structure which suggests that the energy supply may
have been variable. At 92 cm (Fig.~\ref{Fig. 9} bottom) and at 49 cm
(Mack et al.  1997), there is emission south and west of the main radio
axis to the NW. The origin of this emission is unclear.

Double-double radio galaxies (Kaiser et al. \cite{kaiser}) appear to
provide good evidence that activity in AGNs can be recurrent.
Presumably these are objects in which the `on-off' cycles are
comparable. There may be a wide range of duty cycles amongst AGNs;
3C236 may be an example of a source which is occasionally `off',
perhaps due to the jet channel collapsing or the flow being blocked by
interaction with a cloud or a drop in supply of fuel to the central
black hole. The central structure in 3C236 may reflect the normal level
of activity in the nucleus.

Mack et al. (\cite{mack2}) and Schoenmakers et al.
(\cite{schoenmakers}) have estimated the age of 3C236 from synchrotron
loss models and spectral aging as $\sim 10^8$~years. The latter authors
also estimate the speed of advance of the western jet into the
intergalactic medium as 0.1 c. This speed is at the lower limit of what
can be measured with VLBI for compact jets in the nucleus (eg. Owsianik
\& Conway \cite{owsianik}, Tschager et al. \cite{tschager}).

Table 2 lists physical parameters for the core components of radio
source 3C236 calculated assuming equipartition, ellipsoidal geometry
for components B1, B2 and the NW lobe, cylindrical geometry for the
component C, and unity filling factor. The radio window has been taken
from $10^{7} - 10^{11}$~Hz. Component sizes have been measured from the
5 GHz images. The volumes derived for components A and C may be in
error by a factor of 2-3; for B1 and B2 the errors are of the order of
10\%. The spectral indices in the Table are mean values for the
components, with typical errors of 10\%. The radio luminosity and
energy in particles and fields have been calculated following
Pacholczyk (\cite{pacholczyk}), assuming the proton-electron energy
ratio $k=10$. The estimated errors in these quantities vary with the
component in question: for components A and B1 the values listed may be
in error by a factor of 2 to 3, while for components B2 and C the error
estimates are more like 10 to 30 \%. The parameter values found are
comparable with those deduced by Barthel et al. (\cite{barthel}) taking
into account differences in the initial assumptions and measured
component parameters.

\begin{table*}
\caption{Component parameters of 3C236}
\label{table 2}
\begin{center}
\begin{tabular}{lclccc}
\hline \hline
Components & Volume (cm$^{3}$) & Sp index & L$_{radio}$ (erg\,s$^{-1}$) & 
E$_{e}$ (erg)    & E$_{H}$ (erg)        \\[0.1cm] 
\hline
A          & 10$^{62}$         & $-0.4$   & $5\times 10^{41}$           & 
$10^{53}$        & $8\times 10^{54}$    \\[0.1cm]
B2         & $\leq 10^{59}$    & $+0.05$  & $5\times 10^{42}$           & 
$10^{53}$        & $8\times 10^{53}$    \\[0.1cm]
B1         & $10^{59}$         & $-0.6$   & $\ge 2\times10^{41}$        & 
$4\times10^{52}$ & $\ge 3\times10^{53}$ \\[0.1cm] 
C          & $10^{61}$         & $-0.9$   & $4\times 10^{41}$           & 
$7\times10^{53}$ & $6\times 10^{54}$    \\[0.1cm] 
\hline
\hline
\\[1cm]
\end{tabular}
\end{center}
\end{table*}

The jet power measured on large scales by Schoenmakers et al.
(\cite{schoenmakers}) is $\sim 10^{45}$~erg\,s$^{-1}$ assuming an upper
limit of 100 GHz to the radio window. In contrast, on the smaller
linear scales, the jet power is $\sim 5 \times 10^{43}$~erg\,s$^{-1}$
with the same upper limit to the radio emission.  This may indicate
that the nucleus was more active in the past when the outer lobes were
created.

Can we reconcile the symmetric structure seen on $10 - 20$~mas scales
at 1.663 GHz with the one-sided structure seen at 5 GHz on 1 mas
scales?  The HST image (Fig.~\ref{Fig. 9} middle) shows convincingly
that 3C236 is $\sim 30\degr$ from the plane of the sky with the NW jet
approaching. For radio emitting plasma moving along the jets with a
bulk motion less than a few tens of percent of the speed of light, no
beaming effects should be detectable. This is consistent with the
symmetric morphology seen in the nucleus of the galaxy on 10 -- 20 mas
scales (see Fig.~\ref{Fig. 3}). The direction of the one-sided
structure on milli-arcsecond scales is consistent with the NW jet being
the approaching one, but the brightness of the first component in the
jet is  $\sim 50$ times stronger than the noise in the receding jet
direction. This is not consistent with the angle to the line of sight
being 60$\degr$, independent of the magnitude of the bulk velocity
(Baum et al.  \cite{baum}). A boost in intensity of a factor of 50 can
be achieved if the compact jet is oriented at 35$\degr$ to the line of
sight rather than 60$\degr$, and the bulk velocity is $\ge$ 0.8c.  This
implies a substantial change in the orientation of the small scale jet
which is perhaps unlikely in view of the long term stability of the
large scale jet system. However, the P.A. of the compact jet on the
plane of the sky is $10\degr$ different to the large scale P.A.,
indicating that the jet orientation does change.  A more thorough
search for superluminal motion in this structure may be illuminating.

An alternative possibility is that the lack of a counterjet on parsec
scales is caused by free-free absorption by  ionised gas in the inner
region of the gas-dust disk surrounding the black hole.  This is seen
on VLBI scales in NGC 4261 (Jones et al. \cite{jones2}), Cen A (Jones et
al \cite{jones1}), and 3C84 (Walker et al. \cite{walker}; Vermeulen et
al \cite{vermeulen}). In NGC~4261, the 1.6~GHz image is symmetric about
the nucleus as is the case for 3C236, but at 8.4~GHz NGC~4261 shows the
effects of free-free absorption.  Higher frequency observations are
required to test this possibility in 3C236.

\section{Conclusions}

The radio structure in the nucleus of the giant radio galaxy, 3C236,
has been investigated using combinations of VLBI and MERLIN data at
1.663 and 4.987~GHz. The nuclear structure is dominated by two main
complexes of emission whose orientation and separation are in general
well aligned with the large scale structure, although differences in
the projected P.A. reach $\sim 10\degr$ within a few tens of
milli-arcseconds either side of the nuclear component indicating that
the jets oscillate about the mean direction set by the Mpc scale
structure. The nuclear component has been identified by its flatter
spectrum and compact size in spectral index distributions made at
angular resolutions of 10 and 25 mas. The emission is one-sided to the
north-west on scales of a few milli-arcsconds, but symmetric on scales
of a few tens of milli-arcseconds.

Re-examination of HST data on 3C236 taken by de Koff et al.
(\cite{dekoff}) reveals that the normal to the plane of the dust ring
in the centre of the galaxy is at $\sim 60\degr$ to the line of sight
and parallel, within 15$\degr$ to 20$\degr$, to the projected
elongation of the radio structure. Assuming that the radio axis is
parallel to the normal to the dust disk, we identify the shorter
north-west jet as being on the approaching side, and the longer
south-east jet as receding. The true size of 3C236 is therefore 4.5~Mpc
rather than the 3.9 Mpc determined assuming the source is in the plane
of the sky.  Coincidence, within the astrometric errors, of a dust
feature with the compact south-east jet (component C) is unlikely to be
evidence that the south-east radio jet entrains dust in the first
400~pc after leaving the nucleus, rather that the jet has encountered a
cloud of gas and dust which actually shapes the radio component.

The one-sided nuclear radio structure to the north-west can be
explained on the basis of relativistic beaming effects, if the angle to
the line of sight is $\sim 15\degr$ and the bulk motion of the radio
emitting plasma is highly relativistic. However, it is not clear that
such a large deviation in angle to the line of sight compared to the
large scale structure can be reconciled with the relatively small
deviations in projected orientation of the jet system from small to
large scales.  If the structure is due to beaming, then we must
conclude that the change from one-sided to two-sided emission implies
that the angle to the line of sight has reached 60$\degr$ at a
separation of $\sim 20$~mas from the nucleus. The alternative
possibility of free-free absorption of the counter jet needs
investigation.

At the resolution of WSRT data at 327 MHz by Schoenmakers et al.
(\cite{schoenmakers}), the south-east jet appears essentially continuous
over a distance of 2.5 Mpc from the nucleus to the hotspots, making
this the largest jet known in the universe. It is likely, however, that
the central activity is episodic but with a shorter duty cycle than for
the double-double galaxies studied by Schoenmakers et al.  The current
central radio emission in 3C236 is evidence of a recent major burst of
radio emission superimposed on a more uniform level of activity,
perhaps as a result of subsuming a companion galaxy.

\begin{acknowledgements}
We thank A.G.~de~Bruyn for re-mapping the 327 MHz WSRT data by
Schoenmakers et al. (\cite{schoenmakers})  in Fig.~\ref{Fig. 9},
T.W.M.~Muxlow for assistance in combining the VLBI and MERLIN data, and
P.~El\'osegui, H.~Liang and H.~Sanghera for assistance in reducing the
data. We also thank P.~Best for helpful discussions of the HST data,
and M.A.~Garrett for comments on the manuscript. Thanks are also due to
the staff at the EVN and US observatories for carrying out the VLBI
observations, and to the staff at the CIT/JPL and MPIfR correlators for
processing the data. The National Radio Astronomy Observatory is a
facility of the National Science Foundation operated under cooperative
agreement by Associated Universities, Inc.
\end{acknowledgements}

\end{document}